\documentclass[conference]{IEEEtran}
\IEEEoverridecommandlockouts
\usepackage{cite}
\usepackage{amsmath,amssymb,amsfonts}
\usepackage{graphicx}
\usepackage{textcomp}
\usepackage{xcolor}
\usepackage{CJKutf8}
\usepackage{colortbl}
\usepackage{makecell}
\usepackage{hyperref}
\usepackage{balance}
\hypersetup{
    colorlinks=true,
    linkcolor=blue,}
\usepackage{booktabs}
\usepackage{multirow}
\usepackage{algorithm}  
\usepackage{amssymb}
\usepackage{algpseudocode}  
\usepackage[misc]{ifsym}
\usepackage{amsthm}
\usepackage{pifont}
\newcommand{\cmark}{\ding{51}}%
\newcommand{\xmark}{\ding{55}}%
\newtheorem{myDef}{Definition}
\newtheorem{myproblem}{Problem}
\newcommand{\tabincell}[2]{\begin{tabular}{@{}#1@{}}#2\end{tabular}}  
\newcolumntype{P}[1]{>{\centering\arraybackslash}p{#1}}

\def\BibTeX{{\rm B\kern-.05em{\sc i\kern-.025em b}\kern-.08em
    T\kern-.1667em\lower.7ex\hbox{E}\kern-.125emX}}
    
\begin{document}
\begin{CJK}{UTF8}{gbsn}

\title{Learning What You Need from What You Did: Product Taxonomy Expansion with User Behaviors Supervision}

\makeatletter
\newcommand{\linebreakand}{%
  \end{@IEEEauthorhalign}
  \hfill\mbox{}\par
  \mbox{}\hfill\begin{@IEEEauthorhalign}
}
\makeatother

\author{\IEEEauthorblockN{Sijie Cheng\IEEEauthorrefmark{1},
Zhouhong Gu\IEEEauthorrefmark{1},
Bang Liu\IEEEauthorrefmark{6}\IEEEauthorrefmark{3}, 
Rui Xie\IEEEauthorrefmark{4},
Wei Wu\IEEEauthorrefmark{5} and
Yanghua Xiao\IEEEauthorrefmark{1}\IEEEauthorrefmark{2}\textsuperscript{\Letter}}
\IEEEauthorblockA{\IEEEauthorrefmark{1}Shanghai Key Laboratory of Data Science, School of Computer Science, Fudan University, Shanghai, China}
\IEEEauthorblockA{\IEEEauthorrefmark{2}Fudan-Aishu Cognitive Intelligence Joint Research Center, Shanghai, China}
\IEEEauthorblockA{\IEEEauthorrefmark{6}RALI \& Mila, Université de Montréal, Montréal, Québec, Canada}
\IEEEauthorblockA{\IEEEauthorrefmark{3}Canada CIFAR AI Chair}
\IEEEauthorblockA{\IEEEauthorrefmark{4}Meituan, Shanghai, China}
\IEEEauthorblockA{\IEEEauthorrefmark{5}Meituan, Beijing, China}
\IEEEauthorblockA{\{sjcheng20, zhgu20, shawyh\}@fudan.edu.cn}
\IEEEauthorblockA{bang.liu@umontreal.ca, \{rui.xie, wuwei30\}@meituan.com}
}

\maketitle

\begin{abstract}
Taxonomies have been widely used in various domains to underpin numerous applications.
Specially, product taxonomies serve an essential role in the e-commerce domain for the recommendation, browsing, and query understanding.
However, taxonomies need to constantly capture the newly emerged terms or concepts in e-commerce platforms to keep up-to-date, which is expensive and labor-intensive if it relies on manual maintenance and updates. 
Therefore, we target the taxonomy expansion task to attach new concepts to existing taxonomies automatically.
In this paper, we present a self-supervised and user behavior-oriented product taxonomy expansion framework to append new concepts into existing taxonomies.
Our framework extracts hyponymy relations that conform to users' intentions and cognition.
Specifically, i) to fully exploit user behavioral information, we extract candidate hyponymy relations that match user interests from query-click concepts; ii) to enhance the semantic information of new concepts and better detect hyponymy relations, we model concepts and relations through both user-generated content and structural information in existing taxonomies and user click logs, by leveraging Pre-trained Language Models and Graph Neural Network combined with Contrastive Learning; iii) to reduce the cost of dataset construction and overcome data skews, we construct a high-quality and balanced training dataset from existing taxonomy with no supervision.
Extensive experiments on real-world product taxonomies in Meituan Platform, a leading Chinese vertical e-commerce platform to order take-out with more than 70 million daily active users, demonstrate the superiority of our proposed framework over state-of-the-art methods.
Notably, our method enlarges the size of real-world product taxonomies from 39,263 to 94,698 relations with 88\% precision.
Our implementation is available:
\url{https://github.com/AdaCheng/Product_Taxonomy_Expansion}.
\end{abstract}


\section{Introduction}
With the increasing importance of taxonomies in various applications, many taxonomies have been constructed in general and specific domains, such as lexical taxonomy (e.g., WordNet\cite{miller1995wordnet}), scientific taxonomies (e.g., MeSH\cite{lipscomb2000medical}), and product taxonomies built by Amazon\cite{dong2020autoknow}.
Specially, product taxonomies have been widely used as an essential component to underpin various downstream applications in the e-commerce domain, including personalized recommendation\cite{karamanolakis2020txtract, huang2019taxonomy, zhang2014taxonomy}, query understanding \cite{yang2020co, hua2016understand} and web content tagging\cite{liu2019user,liu2020giant,peng2019hierarchical}.
However, as new concepts constantly emerge, existing web-scale product taxonomies face an outdated problem.
Manually maintaining or updating existing product taxonomies is labor-intensive and difficult to scale.
As a result, \emph{the taxonomy expansion task}, which automatically attaches new concepts to the existing taxonomy, has received increasing attention.

Existing relevant methods are mainly designed for general-purpose taxonomies paying great attention to feature extraction from an open domain corpus, having limited effects on product taxonomy. The reasons are as follows.
First, the semantics of domain-specific concepts or new concepts might not be represented well.
The concepts in the e-commerce domain rarely appear in general corpora, making corpus-based methods \cite{shen2018hiexpan, vedula2018enriching, fauceglia2019automatic, aly2019every, mao2020octet} infeasible.
Graph-based methods\cite{manzoor2020expanding, shen2020taxoexpan, yu2020steam, ma2021hyperexpan} can slightly enrich the semantics by aggregating the neighbors' information in the existing taxonomy, but it does not work well for emerging concepts.
Without sufficiently modeling the actual semantic of concepts, it is difficult to infer the hyponymy relations.
Second, existing approaches neglect that e-commerce platforms are highly user-involved systems.
Matching the shopping need of a user is one of the major functionalities of an e-commerce platform\cite{luo2020alicoco}.
Thus, the selection of candidate concepts and the definition of appropriate hyponymy relations rely on user cognition.
Non-user-centered taxonomy expansion introduces a huge gap between user needs and the product taxonomy, which further hurts the utility of taxonomies to serve downstream applications.

The aforementioned considerations motivate us to propose a novel taxonomy expansion method driven by user behavioral data. Such data contains rich effective signals to infer a hypernymy relation, which is the core task in taxonomy expansion. 
Specifically, we mainly use user click logs and user-generated content. 
From the user click logs, we can find many queries and corresponding clicked items. One observation is that the queries in many cases are coarse-grained hypernym concepts, such as \emph{``Bread (面包)''}, while the corresponding clicked items are quite likely fine-grained hyponym products, such as \emph{``Cheese Bun (奶酪包)''}. 
From the user-generated content, we can easily find some user comments such as \emph{``Cheese Bun (奶酪包) is one of my favorite bread (面包)''}, which strongly suggests the hypernymy relation between \emph{``Cheese Bun (奶酪包)''} and \emph{``Bread (面包)''}. 
Therefore, user behavioral data in e-commerce platforms reflects users' search intention and their cognition about items in the platforms.

In addition, hyponymy relation inference in taxonomy expansion generally needs appropriate supervision. Self-supervised strategy which constructs a training dataset automatically from existing taxonomy is popularly used in recent studies\cite{mao2020octet, yu2020steam, shen2020taxoexpan, song2021should, ma2021hyperexpan}.
However, most existing taxonomies are semi-automatically constructed, which inevitably incurs data skews issues.
For example, 96.56\% hyponymy relations in our existing taxonomy (in Section \ref{sec:datasets}) can be detected by headword (e.g., \emph{``Rye Bread (黑麦面包)'' IsA ``Bread (面包)''}). In contrast, the proportion of other difficult cases of hypernymy relations (e.g., \emph{``Cheese Bun (奶酪包)'' IsA ``Bread (面包)''}) is tiny.
Therefore, it is essential to rectify the bias inherent in a web sale taxonomy usually constructed by semi-automatic approaches, ensuring the quality of self-supervised dataset generation.


In this paper, we propose a self-supervised user behavior-oriented product taxonomy expansion framework.
Different from prior work, we not only aim to attach the appropriate concepts to an existing taxonomy, but also further consider conforming to user intention and cognition by mining from user click logs and user-generated content.
To generate labeled data without human effort and avoid error propagation, we utilize adaptive self-supervision to construct a high-quality training dataset constrained by user behaviors, focusing on keeping the appropriate balance among different pattern types.
The most remarkable advantage is that our methods can continuously update the existing taxonomy as user behavior information grows day by day.
Specifically, we make the following contributions in this paper:

\emph{First,} to the best of our knowledge, we propose the first product taxonomy expansion framework driven by user behavioral data. We fuse user click-through and comments into a heterogeneous graph to accomplish the taxonomy expansion task.
The rich user behavioral data helps to prune the search space of candidate hyponymy relations and enrich the representations of concepts in a specific domain.
The heterogeneous graph integrates various types of information from both existing taxonomies and user behavioral data. The heterogeneous graph-based framework is also extensible to incorporate more emerging user data in the e-commerce domain.

\emph{Second,} we propose two effective representation learning strategies that are well tailored for the user behavior driven taxonomy expansion task.
One is implicit relational representations automatically derived from pre-trained language models fine-tuned on user-generated content. The other is structural representations extracted from both exiting taxonomies and user click logs via utilizing contrastive learning and graph neural networks.
Both automatic and manual evaluations have shown that our representation learning method significantly outperforms other baselines.

\emph{Third,} we propose a novel self-supervised dataset construction method that explicitly addresses the data noise and data skew issues inherent in the supervision from a web-scale product taxonomy.
Our solution helps construct high-quality datasets with above 95\% accuracy without human labor. The dataset is also adaptively refined so that the different patterns of samples are subtly balanced, alleviating the over-fitting problem and improving the generalization ability of our hyponymy relation inference model. 

\emph{Last but not least,} we conduct extensive experiments on real large-scale platforms, justifying the effectiveness of the proposed solution. Experimental results show that our proposed method significantly surpasses the previous state-of-the-art baselines on all the datasets by 74.92\% in accuracy and 77.56\% in edge-F1.
We apply our solution to expand the product taxonomy of Meituan (美团)，a leading Chinese vertical e-commerce platform for ordering take-outs with more than 70 million daily active users. Specifically, we significantly enlarge the size (39,263 to 94,698 relations) of Meituan product taxonomies with 88\% precision. Finally, an offline user study shows that query rewriting supported by the expanded taxonomy improves feed stream recommendation by 6\% precision. Although our framework is tested for this specific platform, the core idea to use user behavior data to expand a taxonomy can be easily adapted to any platform rich in user behavioral data.

\begin{figure*}[!t]
    \centering
    \includegraphics[width=1.0\linewidth]{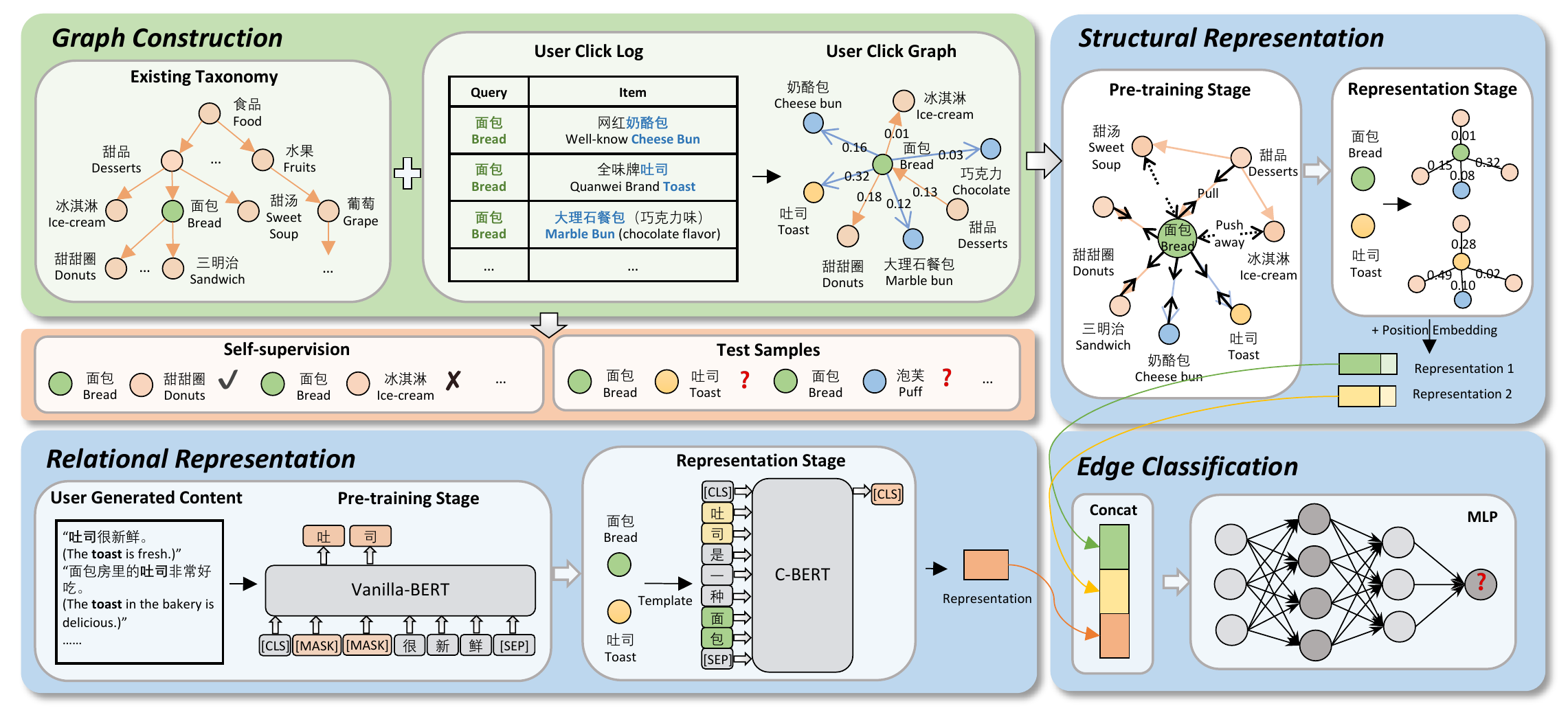}
    \vspace{-7mm}
    \caption{Our proposed framework mainly contains two major phases: \textbf{\textcolor[rgb]{0.1,0.8,0.1}{Graph Construction}} and \textbf{\textcolor[RGB]{30,144,255}{Hyponymy Detection}}. In graph construction module, we build a user click graph combining with existing taxonomy and user click logs. In hyponymy detection module, we firstly obtain the relational representation (bottom-left) and structural representation (top-right), then utilize them to detect the hyponymy relations between a query and its clicked items. In addition, we automatically construct the training and validation datasets via \textbf{\textcolor[RGB]{237,145,33}{adaptively self-supervised generation}}.
    }
    \vspace{-4mm}
    \label{fig:overview}
\end{figure*}

\section{Problem Formulation} \label{sec:problem}
In this section, we first give the definitions of the preliminary concepts and user behavioral data used in this paper, then formally define our taxonomy expansion problems, and further discuss the scope of our study.

\subsection{Preliminaries} \label{sec:define}

\begin{myDef}
\textbf{(Taxonomy.)}
A taxonomy $\mathcal{T}=(\mathcal{N}, \mathcal{E})$ is a tree-structured hierarchy.
Each node $n \in \mathcal{N}$ denotes a concept and each directed edge $\langle n_p, n_c \rangle \in \mathcal{E}$ represents the hyponymy relation between the parent (hypernym) node $n_p$ and the child (hyponym) node $n_c$.
\end{myDef}

\begin{myDef}
\textbf{(Concept Vocabulary.)}
To assure the quality of candidate concepts extracted from user behavioral data, we firstly automatically collect a set of product concepts $\mathcal{P} = \{p_1, p_2, \cdots, p_k\}$ from menus in merchants. Then we ask experts to manually label them and finally obtain a clean product concept vocabulary $\mathcal{C} = \{c_1, c_2, \cdots, c_m\}$.
\end{myDef}

\begin{myDef}
\textbf{(User Click Logs.)}
User click logs $\mathcal{U}_{l} = \{(q, I_q)\}$ record each user query $q\in \mathcal{Q}$ and its corresponding clicked items $I_q\in \mathcal{I}$. 
Each item in ${I}_q$ is represented by its descriptive text, which is usually defined by the merchant, such as ``Well-known Cheese Bun (网红奶酪包)''.
\end{myDef}

User click logs reflect user search intention and contain rich information about hyponymy relations. For example, when we query \emph{``Bread (面包)''}, \emph{``Cheese Bun (奶酪包)''} is one of the typical clicked items. Such search and click behavior reveals that \emph{``Cheese Bun (奶酪包)''} can be seen as an instance or a sub-concept of \emph{``Bread (面包)''}. Furthermore, as a hypernym of \emph{``Cheese Bun (奶酪包)''}, \emph{``Bread (面包)''} also reveals the user's real search intention.

\begin{myDef}
\textbf{(User-generated Content.)}
User-generated content $\mathcal{U}_{c}=\{t_1, t_2, \cdots, t_n\}$ refers in particular to the corpus of $n$ texts generated by users, such as the reviews of products.
\end{myDef}

User-generated content (UGC) directly expresses user cognition of items and contains vast implicit domain-specific relational knowledge.
Suppose we are given \emph{``Bread (面包)''} and \emph{``Toast (吐司)''} to predict whether there exists hyponymy relation between them. 
UGC might implicitly mention their relations.
For example, it is pretty likely to find two sentences \emph{``The toast in this bakery is delicious (这家面包店的吐司真美味)''} and \emph{``The bakery sells all kinds of bread (面包店里卖各种各样的面包)''} from the whole UGC, which are enough to infer that \emph{``Toast (吐司)''} is a kind of \emph{``Bread (面包)''}.

\subsection{Problem Definition}
\begin{myproblem}
\textbf{(Taxonomy Expansion.)}
Given an existing taxonomy $\mathcal{T}^{0}=(\mathcal{N}^{0}, \mathcal{E}^{0})$, the set of clean concept vocabulary $\mathcal{C}$, the goal of taxonomy expansion task is to attach the appropriate concept $c \in \mathcal{C}$ to the existing taxonomy $\mathcal{T}^{0}$ and expand it to obtain an enriched taxonomy $\mathcal{T}^{*}=(\mathcal{N}^{0} \cup \mathcal{C}, \mathcal{E}^{0} \cup \mathcal{R})$, where $\mathcal{R}$ is a set of new hyponymy relations $\langle n, c \rangle$, $n \in \mathcal{N}^{0}$, $c \in \mathcal{C}$.
\end{myproblem}

Formally, we consider each node $n \in \mathcal{N}^{0}$ as a random variable and the taxonomy $\mathcal{T}^{0}$ as a Bayesian network.
The probability of taxonomy $\mathcal{T}^{0}$ can be formulated as follows.
\begin{equation}
\small
\mathbf{P}(\mathcal{T}^{0} \mid \Theta)=\mathbf{P}(\mathcal{N}^{0} \mid \mathcal{T}^{0}; \Theta)=\prod_{i=1}^{|\mathcal{N}^{0}|} \mathbf{P}\left(n_{i} \mid p\left(n_{i}\right); \Theta\right).
\end{equation}
where $\Theta$ denotes model parameters, and $p\left(n_{i}\right)$ represents the parent node(s) of $n_{i}$.
We then find the optimal taxonomy $\mathcal{T}^{*}$ by maximizing the likelihood.
\begin{equation}
\small
\begin{aligned}
\mathcal{T}^{*} &=\underset{\mathcal{T}^{0}}{\arg \max } \mathbf{P}(\mathcal{T}^{0} \mid \Theta) \\
&=\underset{\mathcal{T}^{0}}{\arg \max } \sum_{i=1}^{\left|\mathcal{N}^{0} \cup \mathcal{C}\right|} \log \mathbf{P}\left(n_{i} \mid p\left(n_{i}\right); \Theta\right).
\end{aligned}
\end{equation}

However, for every node in the existing taxonomy, all concepts in $\mathcal{C}$ can be taken as candidates of hyponym, thus the potential search space is incredibly enormous.
Previous studies \cite{manzoor2020expanding, shen2020taxoexpan, yu2020steam, zhang2021taxonomy, mao2020octet} address the above problem via an assumption that the input set of new concepts contains only one element (i.e., $|\mathcal{C}|=1$), and aims to find one single parent node in the existing taxonomy $\mathcal{T}^{0}$ of this new concept (i.e., $|\mathcal{R}|=1$).
This simplification not only limits expansion in the depth of taxonomies, but also neglects the truth that a new concept has multiple parent nodes.
In this paper, we discard this assumption and propose restricting this search space through strong prior hyponymy relations in user behaviors.

\section{Solutions} \label{sec:solution}
In this section, we propose an adaptively self-supervised user-behavior oriented framework for product taxonomy expansion as illustrated in Figure \ref{fig:overview}.
In our solutions, we model the taxonomy expansion as an \emph{edge classification problem}, which is different from previous methods \cite{shen2020taxoexpan, yu2020steam} that model taxonomy expansion task as a multi-class ranking problem or a shortest path finding problem.
We first elaborate on our two key modules: graph construction and hyponymy detection, then discuss the process of model learning and inference.

\subsection{Graph Construction} \label{sec:graph}

Given the set of clean concept vocabulary $\mathcal{C}$, an existing taxonomy $\mathcal{T}^{0}$ and user click logs $\mathcal{U}_{l}$, here we need to construct a heterogeneous edge-weighted graph $\mathcal{G}_h = (\mathcal{N}_{h}, \mathcal{E}_{h})$ in the following four steps.

\subsubsection{Items Collection} 
Based on the existing taxonomy, we collect items through treating concepts in existing taxonomy as query concepts $\mathcal{C}_q$ to find their clicked items. For example, there is a concept \emph{``Bread (面包)''} in the existing taxonomy. If it is a query concept that appeared in the user click logs, its corresponding clicked items, such as \emph{``Well-known Cheese Bun (网红奶酪包)''}, are used as candidate nodes in the graph. 

\subsubsection{Nodes Identification}
As an item can be described by different names, we identify the clicked concept nodes $\mathcal{C}_i$ in the clean concept vocabulary $\mathcal{C}$ via longest common sub-string matching.
For example, given the set of concept vocabulary $\mathcal{C}=\{\emph{``Bun (包)''},\ \emph{Cheese Bun (奶酪包)''},\ \cdots \}$, the concept node \emph{``Cheese Bun (奶酪包)''} is identified from the clicked item \emph{``Well-known Cheese Bun (网红奶酪包)''}.

\subsubsection{Edge Connection}
Next, we connect query concepts $\mathcal{C}_q$ and item concepts $\mathcal{C}_i$, representing their interactions in user behavioral data.
In this way, we can acquire a new edge between \emph{``Bread (面包)''} and \emph{``Cheese Bun (奶酪包)''}.

\subsubsection{Weight Assignment}
Considering that the probabilities of different edges to be hyponymy relations should not be equal, and user click behavior can reflect confidences among edges, we use a variant of the Term Frequency - Inverse Document Frequency\cite{ramos2003using} to set our edge weights.

Our Item Frequency (IF) denotes the clicked frequency of the item concept $c_i \in \mathcal{C}_{i}$ according to the query concept $c_q \in \mathcal{C}_{q}$, normalized by the total clicked frequencies of item concepts under the query concept. Intuitively, under the same query concept, IF indicates the importance of item concepts where the item concept with more clicked times is more likely to be the hyponym.
\begin{equation}
    \small 
    \text{IF}_{c_q, c_i} = \frac{{num}_{c_q, c_i}}{\underset{c_k\in \mathcal{C}_i}{\sum}{num}_{c_q, c_k}}.
\end{equation}

Our Inverse Query Frequency (IQF) denotes the total number of query concepts $\mathcal{C}_q$ in the user click graph, divided by the total number of query concepts $c_q$ which click to the item concept $c_i$. 
\begin{equation}
    \small
    \text{IQF}_{c_i} = \log\frac{|\mathcal{C}_q|}{|\left\{c_q:c_q \rightarrow c_i\right\}|}.
\end{equation}

Comprehensively considering importance and novelty factors, the attribute $a_{c_q,c_i}$ of edge $e_{c_q,c_i} = \langle c_q, c_i \rangle$ can be calculated by multiplying the two scores together.
We adopt a square to balance these two factors and further normalize the attribute score to ensure it is between 0 and 1, while the sum of weights under the same query concept is 1. Meanwhile, the weights of edges from the existing taxonomy are all set to 1.

\begin{equation}
    \small
    a_{c_q,c_i} = \text{softmax}(\text{IF}_{c_q, c_i} * (\text{IQF}_{c_i})^{2}).
\end{equation}

Through this algorithm of weight assignment, we can also alleviate two types of noises in the user behavioral data:

(i) \textbf{Intention-drifted behavior: } 
People click the items which are not consistent with their initial intention, which leads to semantic drift.
For example, some people intend to buy \emph{``Bread (面包)''} at first.
However, there are some distractors such as \emph{``Sweet Soup (甜汤)''} shown in the browsing page, because both \emph{``Bread (面包)''} and \emph{``Sweet Soup (甜汤)''} belong to \emph{``Dessert (甜点)''}.
Due to these distractors, several of them end up buying \emph{``Sweet Soup (甜汤)''} which does not have hyponymy relation with \emph{``Bread (面包)''}. 
With the critical factor IF, this type of noise will be assigned a lower weight.

(ii) \textbf{Common but non-sense behavior:} Some common concepts that occurred in the click items to all queries without hyponymy relations.
For instance, some Chinese users love to drink soup during mealtime, so they may order an extra \emph{``Sweet Soup (甜汤)''} no matter what they intend to order.
However, \emph{``Sweet Soup (甜汤)''} is not a suitable hyponymy item concept for most query concepts, such as \emph{``Bread (面包)''}.
The weight of common items that appeared under most query concepts will be punished through the novelty factor IQF.

\subsection{Hyponymy Detection}
We propose a hyponymy detection module to classify the edges in the heterogeneous graph.
Unlike previous methods that manually design a set of shallow lexical or structural features\cite{mao2020octet, shen2020taxoexpan}, we automatically capture two complementary representations of concepts via user behavioral information.

\subsubsection{Relational Representation from User-Generated Content}
User-Generated Content, e.g., user reviews or comments on a platform, are the main text corpora in the e-commerce domain.
They contain rich but noisy context information of domain-specific concepts. 
The hyponymy relations between concepts are usually implicitly expressed in UGC instead of explicitly stated.
Therefore, directly extracting the exact hyponymy relations via pattern-based methods, such as Hearst patterns\cite{hearst1992automatic, roller2018hearst}, is infeasible.
To solve this problem, we propose using Pre-trained Language Models to capture relational knowledge from UGC automatically.

\textbf{Pretraining Stage.}
Due to the lack of domain knowledge in vanilla BERT-Chinese \cite{cui2019pre} that pretrained on general text corpora, we first pre-train a BERT-Chinese on User Generated Content. 
Considering that our target is to classify the hyponymy relation between two concepts, we adopt a concept-level masking strategy to better model the representations of concepts rather than following the token-level masking strategy in the vanilla BERT-Chinese.
As shown in Figure \ref{fig:overview}, in the pre-training stage of relational representation, we first employ our internal word segmentation tool (which can be replaced by other off-the-shelf text segmentation tools such as Jieba) to segment the sentence. Then we select nouns and match them to the given concept vocabulary.
After detecting all the concepts in the sentence, we randomly mask the concepts mentioned in the sentence with \texttt{[MASK]} token and ask the model to recover all slots.
Finally, we acquire an enhanced model called C(oncept)-BERT, modeling domain-specific concept knowledge.

\textbf{Representation Stage.}
Traditional expansion methods always concatenate the representations of two concepts or compute their similarity score.
We argue that there exist several disadvantages:
first, the representations only contain semantic information of each concept but neglect the relational information between them;
second, the direction in the hyponymy relations are not reflected in the representations;
third, empirical experiments have confirmed that BERT performs better in sentence-level representation than token-level and phrase-level representation.

To this end, we manually design a pre-defined template to capture the relational representation following \cite{chen2020inducing}.
Given a query concept $c_q$ and an item concept $c_i$, the input format to C-BERT is organized as below.
\begin{equation}
    \small
    \text{Input}_{c_q, c_i}=[\langle\mathrm{CLS}\rangle \oplus c_q \oplus is \oplus a \oplus c_i \oplus \langle\mathrm{SEP}\rangle].
\end{equation}
where $\oplus$ represents concatenation, $\langle\mathrm{CLS}\rangle$ and $\langle\mathrm{SEP}\rangle$ are the special tokens for classification and sentence separation respectively.
We then encode $D_{n_p, n_c}$ by C-BERT(·), and take the final layer representation of $\langle\mathrm{CLS}\rangle$ as the relational representation of the concept pair $\langle c_q, c_i \rangle$.
\begin{equation}
    \small
    r_{c_q, c_i}=\text{C-BERT}(\text{Input}_{c_q, c_i})[0].
\end{equation}

\subsubsection{Structural Representation from Heterogeneous Graph}
Given the weighted heterogeneous graph $\mathcal{G}_h = (\mathcal{N}_{h}, \mathcal{E}^{0})$, our goal is to capture the structural representation from both existing taxonomy and user click graph for each node.
To achieve that, we leverage Graph Neural Networks combined with Contrastive Learning to finetune the whole graph and then represent the structural information.

\textbf{Pretraining Stage.}  
Since there exist strong hyponymy relations between neighbor node pairs and other node pairs are likely to be unrelated,
inspired by the recent success of contrastive learning in graph learning \cite{qiu2020gcc, you2020graph}, we propose to adopt contrastive learning as our pre-training task to pull together the neighbor node pairs and push away others.

For each node $u \in \mathcal{N}_{h}$, the initial input representation $h_{u}^0$ of $u$ is produced by C-BERT.
\begin{equation}
    \small
    h_{u}^0 = \text{C-BERT}(\langle\mathrm{CLS}\rangle \oplus u \oplus \langle\mathrm{SEP}\rangle)[0].
\end{equation}

We first compute the cosine similarity between the hidden representations of two nodes $u$ and $v$.
\begin{equation}
    \small
    S(u,v) = \frac{h_{u}^{0} h_{v}^{0}}{\parallel {h_{u}^{0} \parallel\parallel h_{v}^{0}} \parallel},
\end{equation}
where $\parallel\cdot\parallel$ is $L_2$-norm.
Then we set node $u \in \mathcal{N}_{h}$ as an anchor node and define its neighbors as $N(u)$.
The node pairs between $u$ and neighbors $N(u)$ are positive samples, and the node pairs between $u$ and all the other nodes in the heterogeneous graph are negative samples.
For an anchor node $u$, the loss function is defined based on InfoNCE\cite{oord2018representation} as follows:
\begin{equation}
    \small
    \mathcal{L}_{CL} = -log \frac{\sum_{v \in N(u)}exp(S(u,v))}{\sum_{v \in \mathcal{N}_h}exp(S(u,v))}.
\end{equation}

\textbf{Representation Stage.}
It is natural to leverage graph neural networks to update the representations of $u$ iteratively $\in \mathcal{N}_{h}$ by aggregating the representations of its neighbors $N(u)$ and itself in a heterogeneous graph $\mathcal{G}_h$.
For convenience, we define the neighbors $N(u)$ and itself $u$ as $\widetilde{N(u)}$.
The structural representation updates information within its K-hop neighborhood after $K$ iterations.
We formulate the hidden representation of $u$ with K-layer(hop).
\begin{equation}
    \small
    h_{u}^{k}=\text{AGG}^{k}\left(\left\{h_{v}^{k-1} \mid v \in \widetilde{N(u)}\right\}\right), k \in\{1, \ldots, K\},
\end{equation}
where the $\text{AGG}^{k}$ is an aggregation function in the k-th layer which we expressly adopt Graph Convolutional Network (GCN)\cite{kipf2016semi} as follows.
\begin{equation}
    \small
    h_{u}^{k}=\rho\left(\sum_{v \in \widetilde{N(u)}} a_{u v}^{k-1} \mathbf{W}^{(k-1)} h_{v}^{(k-1)}\right),
\end{equation}
where $a_{u v}^{k-1}$ is the edge attribute in the heterogeneous graph between node $u$ and node $v$ which is same for all layers, $\rho(\cdot)$ denotes an activation function (e.g., ReLU), and $\mathbf{W}^{(k-1)}$ is the learnable weight matrix.

A significant flaw of the above GNN model is that it does not consider the direction of the hyponymy relation between two concepts.
To solve this problem, given a query concept $c_q$ and an item concept $c_i$, we concatenate a position embedding $p_\text{parent}$ to the former, and $p_\text{child}$ to the later node.
These two types of position embedding, $p_\text{parent}$ and $p_\text{child}$, indicate the node identity of the parent or child.
We define the final structural representation of a node pair $(c_q, c_i)$ as follows.

\begin{equation}
    \small
    s_{c_q, c_i} = [h_{c_q}^{k} \oplus p_\text{parent} \oplus h_{c_i}^{k} \oplus p_\text{child}].
\end{equation}

\subsubsection{Edge Classification}
For each edge $\langle c_q, c_i \rangle$ between a query concept $c_q$ and an item concept $c_i$, we concatenate the representations detailed above to the edge representation:
\begin{equation}
    \small
    e_{c_q, c_i} = [r_{c_q, c_i} \oplus s_{c_q, c_i}].
\end{equation}

We then feed the edge representation $e_{c_q, c_i}$ into a multi-layer perception with one hidden layer to classify whether there is a hyponymy relation between a query concept $c_q$ and an item concept $c_i$.
\begin{equation}
    \small
    f^{\text{MLP}}(c_q, c_i) = \gamma(W_2\sigma(W_1e_{c_q, c_i}+B_1)+B_2),
\end{equation}
where $\{W_1, B_1, W_2, B_2\}$ are parameters; $\gamma(\cdot)$ is the \texttt{Softmax} function, and $\sigma(\cdot)$ is the \texttt{Sigmoid} activation function.

\begin{figure}[!t]
    \centering
    \includegraphics[width=1.0\linewidth]{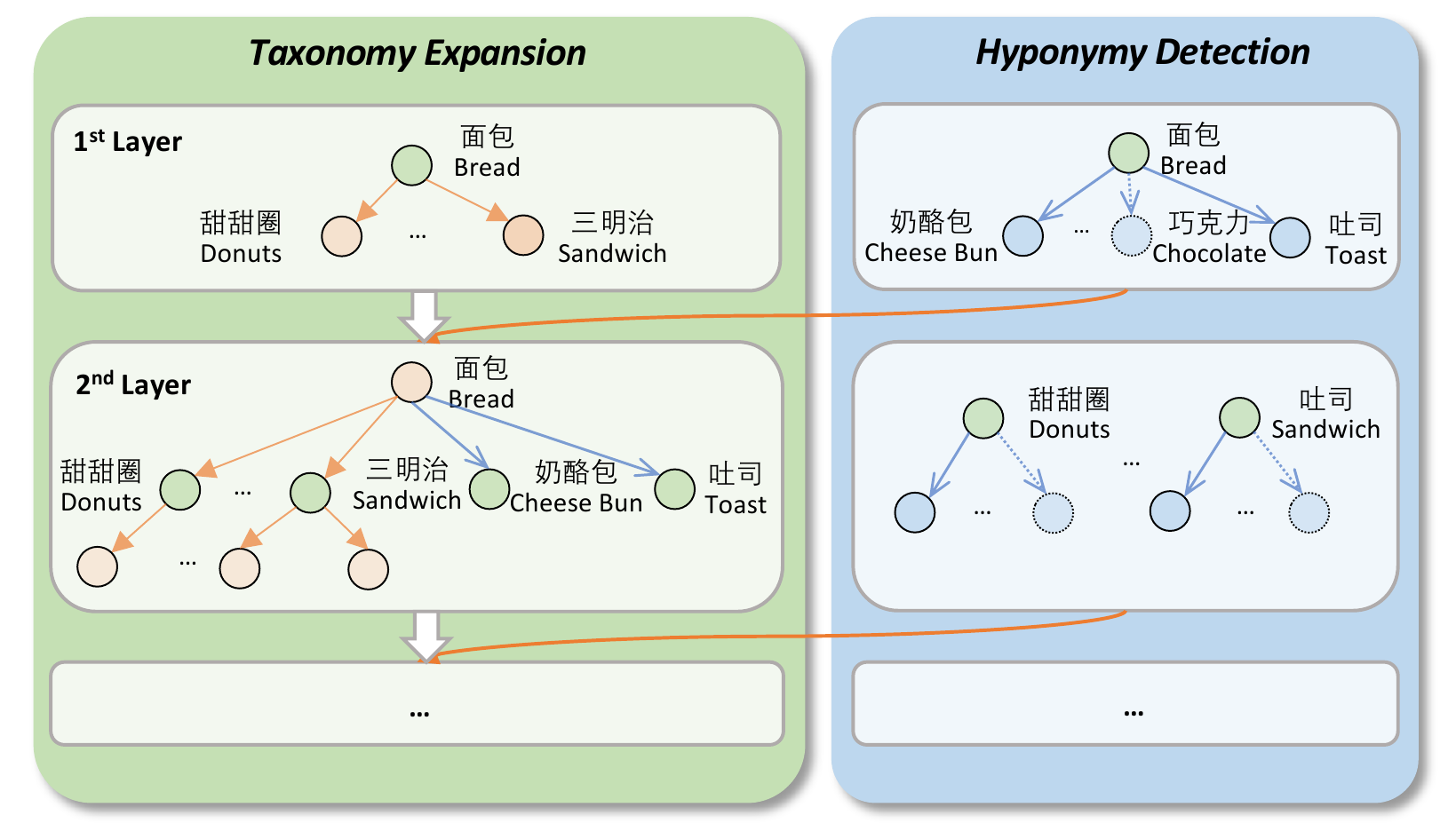}
    \vspace{-6mm}
    \caption{Top-down strategy for updating the existing taxonomy.}
    \vspace{-3mm}
    \label{fig:inference}
\end{figure}

\subsection{Model Learning and Inference} \label{sec:inference}
To learn the hyponymy detection module, we first introduce the self-supervised generation process using edges from existing taxonomies, and then discuss how to perform model training and inference.

\subsubsection{Self-supervised Data Generation}
We prefer to generate self-supervision data from existing taxonomy $\mathcal{T}^{0}=(\mathcal{N}^{0}, \mathcal{E}^{0})$ with no human curation involved.
However, there is a critical challenge that existing taxonomy is exceptionally unbalanced in terms of hyponymy relations: about 93\% of the hyponymy relations can be detected by headword, but other hyponymy relations are rarely contained.
For example, “xxx Bread” is a kind of \emph{“Bread (面包)”}, while some concepts such as \emph{“Toast (吐司)”} and \emph{“Baguette (长棍面包)”} is also the hyponymy concepts of \emph{“Bread (面包)”}.
Nevertheless, the set of hyponymy concepts under the concept \emph{“Bread (面包)”} are almost \{\emph{“Red Bean Bread (红豆面包)”}, \emph{“Fried Bread (油炸面包)”}, \dots\} in the existing taxonomy.
We need to extract more hyponymy relations of diverse patterns that headwords or pattern-based methods can not easily detect.

\textbf{Positive Samples.}
To construct a new balanced supervision hyponymy relations data, we sample hyponymy relations that could not be detected with headword patterns. Specifically, for a specific $\langle c_p, c_i \rangle$ hyponymy relation in existing taxonomy $\mathcal{T}^{0}$, if it can not be detected with headword, it is kept in the training data.
Otherwise, it is selected into the dataset with a probability when the hyponymy reaction appears in the user click data. Through this way, we finally obtain a proportionally balanced taxonomy $\widetilde{\mathcal{T}^{0}}=(\mathcal{N}^{0}, \widetilde{\mathcal{E}^{0}})$.

\textbf{Negative Samples.}
Given an edge $\langle c_q, c_i \rangle$ in the filtered taxonomy as a positive sample, we construct $N$ negative samples alternatively with the following strategy:
\begin{itemize}
    \item Shuffle the order of the edge $c_i, c_q$;
    \item Fix the query concept $c_q$ and replace the item concepts by sampling $N$ concepts $\{c_{n}^{1}, ..., c_{n}^{N}\}$ from user click logs, which are nodes in the filtered taxonomy but neither parents nor descendants of $c_q$.
\end{itemize}

In this way, we collect training instances $X = \{\langle c_q, c_i \rangle, \langle c_i, c_q \rangle, \langle c_q, c_{n}^{1} \rangle, ..., \langle c_q, c_{n}^{N} \rangle\}$ and repeat the process for each edge in the filtered taxonomy $\widetilde{\mathcal{T}^{0}}$.
Our generated full self-supervision training data is $\mathbb{X} = \{X_1, X_2, ..., X_{|\widetilde{\mathcal{E}^{0}|}}\}$.

\subsubsection{Model Training}
At the training stage of edge classification, we learn our representations and adopt Binary Cross-Entropy loss as the objective function of training data $\mathbb{X}$.
\begin{equation}
    \small
    \mathcal{L} = \sum_{\langle c_q, c_i \rangle \in \mathbb{X}}\text{BCELoss}(f^{\text{MLP}}(c_q, c_i), \hat{f}^{\text{MLP}}(c_q, c_i)).
\end{equation}
where $f^{\text{MLP}}(c_q, c_i)$ is the predicted label and $\hat{f}^{\text{MLP}}(c_q, c_i)$ is the ground truth label.

\begin{table*}[!t]
    \caption{Statistics of Term Extraction.}
    \label{tab:extraction}
    \centering
    \begin{tabular}{c|c|ccccc|ccc|c}
        \toprule
        \multirow{2}{*}{\textbf{Taxonomy}} & \multirow{2}{*}{\textbf{\#Items}} & \multicolumn{5}{c|}{\textbf{Existing Taxonomy}} & \multicolumn{3}{c|}{\textbf{New Concept}} & \multirow{2}{*}{\textbf{\#IOthers}} \\
         & & \#Nodes & CNode & \#IEdge & \#Edges & CEdge & \#Concepts & \#INewEdge & \#NewEdge &  \\
        \midrule
         Snack & 60628252 & 19422 & 65.27 & 2021148 & 16319 & 54.15 & 18956 & 15492911 & 12094952 & 43114193 \\
         Fruit & 20368790 & 3032 & 62.43 & 217309 & 2563 & 51.74 & 2639 & 8802659 & 3970631 & 11348822\\
         Prepared Food & 17833703 & 2835 & 68.56 & 207284 & 2498 & 59.83 & 2130 & 8718102 & 1702385 & 8908317\\
        \bottomrule
    \end{tabular}
    \vspace{-5mm}
\end{table*}

\begin{table}[!t]
    \caption{Taxonomy Statistics. $|\mathcal{D}|$ indicates the depth of each taxonomy.}
    \label{tab:Taxonomy}
    \centering
    \begin{tabular}{cccccc}
    \toprule
        \textbf{Taxonomy} & \textbf{$|\mathcal{D}|$} & \textbf{$|\mathcal{N}|$} & \textbf{$|\mathcal{E}|$} & \textbf{$|\mathcal{E}_\text{Head}|$} & \textbf{$|\mathcal{E}_\text{Others}|$} \\
    \midrule
        Overall & 16 & 352940 & 381010 & 330902 & 50308  \\
        Snack & 12 & 29758 & 30134 & 28580 & 3626  \\
        Fruits & 6 & 4857 & 4954 & 4422 & 1241  \\
        Prepared Food & 7 & 4135 & 4175 & 3586 & 1374 \\
    \bottomrule
    \end{tabular}
    \vspace{-4mm}
\end{table}

\subsubsection{Model Inference}
At the inference stage, to expand the existing taxonomy with hyponymy relations between new concepts rather than attach new concepts to existing nodes, we especially update the existing taxonomy by adopting a top-down strategy shown in Figure \ref{fig:inference}.
We traverse the existing taxonomy in level-order.
We apply the learned model for each node in the existing taxonomy to classify and attach the predicted hyponymy relation to the existing taxonomy.
Considering the transitive property of taxonomy\cite{sang2007extracting}, we prune the expanded taxonomy to assure that there is no redundant edge that can infer from the path.
The attached new nodes are also considered for further expanse when we process the next layer.
Through this strategy, we traverse the existing taxonomy only once and expand the width as well as the depth.

\section{Experiments} \label{sec:experiments}
In this section, we first do a series of statistics in the graph construction phase to test whether the hypothesis of strong hyponymy relations exists for query-item interactions in user behavior information.
Second, we compare our proposed approach with various baseline methods.
Subsequently, we further analyze our performance through ablation and case studies.
Finally, we deploy the extended taxonomy in a production environment to perform large-scale online A/B tests.

\subsection{Evaluation on Term Extraction} \label{sec:evaluation}

\subsubsection{Datasets} \label{sec:datasets}
Meituan platform is a leading Chinese vertical e-commerce platform that mainly serves for ordering take-out.
In this paper, we take the Meituan Gourmet Food Taxonomy as the major testbed for all the following experiments.
The depth of Meituan Gourmet Food Taxonomy is 16 layers.
It contains about 195 thousand concepts and 585 thousand taxonomic relations.
We sample three different domain taxonomies from Meituan Gourmet Food taxonomy: Snack, Fruits, and Prepared Food.
The detailed statistics of taxonomies are shown in Table \ref{tab:Taxonomy}. The taxonomy of Snack is 12 layers which is deeper than Fruits (6 layers) and Prepared Food (7 layers). In all taxonomies, the proportion of edges detected by headword is much more significant than others.

\subsubsection{Metrics}
We design a set of metrics to prove the validity of the potential strong  hyponymy relations in user click logs.
\begin{itemize}
    \item \textbf{\#Items} denotes how many query-item records are extracted from user click logs given nodes as query concepts in the existing taxonomy.
    \item \textbf{\#Nodes} denotes how many nodes in the existing taxonomy have their clicked items in user click logs.
    \item \textbf{CNode} calculates the proportion of \#Nodes to the $|\mathcal{N}|$ in the existing taxonomy.
    \item \textbf{\#IEdge} denotes how many query-item concept pairs contain the hyponymy relations in the existing taxonomy.
    \item \textbf{\#Edges} denotes how many hyponymy relations in the existing taxonomy emerged in user click logs as a query-item concept pair.
    \item \textbf{CEdge} calculates the proportion of \#Edges to the $|\mathcal{E}|$ in the existing taxonomy.
    \item \textbf{\#Concepts} denotes how many new concepts are contained in clicked items.
    \item \textbf{\#INewEdge} denotes how many query-item concept pairs can extract new potential hyponymy relations.
    \item \textbf{\#NewEdge} denotes how many new query-item concept pairs can be extracted from user click logs.
    \item \textbf{\#IOthers} denotes how many items contain neither nodes in existing taxonomy $\mathcal{T}$ and concepts in concepts set $\mathcal{C}$. 
    \item \textbf{\#CoNewEdge} calculates the count of correct hyponymy relations in extracted potential hyponymy relations. 
    \item \textbf{Accuracy} calculates the proportion of the query-item pairs which have hyponymy relations. 
\end{itemize}

\subsubsection{Implementation Details}
Given the user click logs and 448,191 manually labeled concepts on the Meituan platform from January 1, 2021, to June 1, 2021, we first automatically compute a set of metrics to demonstrate a large number of potential strong hyponymy relations in user click logs.
We further identify the reasons for not covering other nodes with hyponymy relations.
In terms of accuracy, we randomly sample 10 or 20 query concepts and obtain their corresponding clicked item concepts.
Specifically, we take three taxonomies to manually identify the accuracy of hyponymy relationships in query-item concept pairs.

\begin{table*}[!t]
    \caption{Self-supervised Generated Datasets Statistics. $|\mathcal{E}|$ indicates the total number of the set $\mathcal{E}$.}
    \label{tab:dataset}
    \centering
    \begin{tabular}{ccccccccccc}
    \toprule
        \multirow{2}{*}{\textbf{Dataset}} & \multirow{2}{*}{\textbf{$|\mathcal{E}_\text{All}|$}} & \multicolumn{2}{c}{\textbf{$|\mathcal{E}_\text{All}|$}} & \multicolumn{2}{c}{\textbf{$|\mathcal{E}_\text{Positive}|$}} &  \multicolumn{2}{c}{\textbf{$|\mathcal{E}_\text{Negative}|$}} & \multicolumn{3}{c}{\textbf{$|\mathcal{E}_\text{All}|$}} \\
        \cmidrule(lr){3-4}
        \cmidrule(lr){5-6}
        \cmidrule(lr){7-8}
        \cmidrule(lr){9-11}
        & & \textbf{$|\mathcal{E}_\text{Positive}|$} & \textbf{$|\mathcal{E}_\text{Negative}|$} & \textbf{$|\mathcal{E}_\text{Head}|$} & \textbf{$|\mathcal{E}_\text{Others}|$} & \textbf{$|\mathcal{E}_\text{Shuffle}|$} & \textbf{$|\mathcal{E}_\text{Replace}|$} & \textbf{$|\mathcal{E}_\text{Train}|$} & \textbf{$|\mathcal{E}_\text{Val}|$} & \textbf{$|\mathcal{E}_\text{Test}|$}\\
    \midrule
        Snack & 10360 & 5180 & 5180 & 1554 & 3626 & 2599 & 2581 & 6216 & 2072 & 2072\\
        Fruits & 3546 & 1773 & 1773 & 532 & 1241 & 885 & 888 & 2127 & 709 & 710 \\
        Prepared Food & 3926 & 1963 & 1963 & 589 & 1374 & 975 & 988 & 2355 & 785 & 786\\
    \bottomrule
    \end{tabular}
    \vspace{-5mm}
\end{table*}

\subsubsection{Evaluation Results}
Table \ref{tab:extraction} shows the term extraction statistics.
There are tens of millions of query-item concept pairs in six months of user click records.
We can see that the average coverage of nodes and edges in the existing taxonomy is 64.11\% and 57.66\%, respectively, indicating that user click logs contain rich hyponymy relationships.
Also, it is worth noting that our user click logs are continuously growing, and thus the number of all potential hyponymy relations in the user click logs is much larger.

For the uncovered nodes, we take the Snack domain as an example. We find that 77.00\% of them are leaf nodes, and users are not interested in 18\% (no query logs). Other domains have similar distributions as shown in Figure \ref{fig:uncovered}.
This phenomenon is reasonable because most leaf nodes tend to have no hyponym.
In addition, the part of concepts that users do not ask for is not urgent in our objectives, and our approaches will take them into consideration once they appear in the user click logs.

\begin{figure}[!t]
    \centering
    \vspace{-2mm}
    \includegraphics[width=0.48\textwidth]{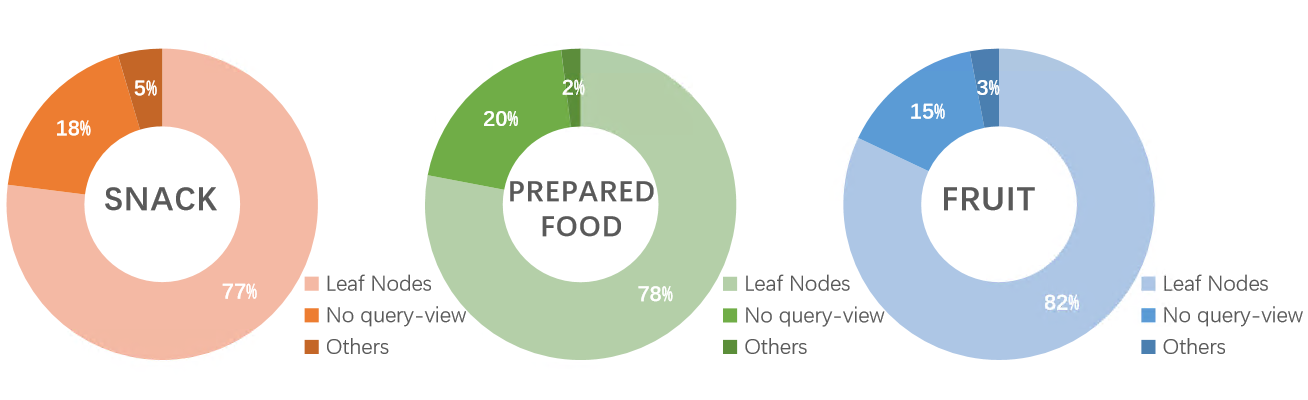}
    \vspace{-4mm}
    \caption{Analysis of the proportion of uncovered nodes in user click logs in Snack domain.}
    \vspace{-5mm}
    \label{fig:uncovered}
\end{figure}

Given a set of concepts $\mathcal{C}$, the number of new extracted concepts in Table \ref{tab:extraction} is nearly consistent with the nodes in the existing taxonomy, and the number of new extracted edges is 10 times larger than the edges, both show that our upper bound of potential search space to expand taxonomy is sufficient.
Furthermore, there still remains multi-million items that may contain potential concepts that have not yet been explored.
Because the high quality and numerous concept set are provided by merchants, we first try to attach these concepts to the existing taxonomy and leave automatically extracting concepts from user click logs in the future.

\begin{table}[!t]
    \caption{Accuracy of Term Extraction.}
    \label{tab:extraction_human}
    \centering
    \begin{tabular}{cccccccc}
        \toprule
        \textbf{Taxonomy} & \textbf{\#Nodes} & \textbf{\#NewEdge} & \textbf{Accuracy}\\
        \midrule
         Snack & 20 & 6986 & 13.18  \\
         Fruits & 10 & 10129 & \textcolor{white}{0}8.46\\
         Prepared Food & 10 & 10212 & 13.00\\  
        \bottomrule
    \end{tabular}
    \vspace{-5mm}
\end{table}

\begin{table*}[!t]
    \caption{Automatic Evaluation.}
    \label{tab:compare}
    \centering
    \begin{tabular}{c|ccc|ccc|ccc}
        \toprule
        \multirow{2}{*}{\textbf{Method}} & \multicolumn{3}{c|}{\textbf{Snack}} & \multicolumn{3}{c|}{\textbf{Fruits}} & \multicolumn{3}{c}{\textbf{Prepared Food}} \\
         & Acc & Edge-F1 & Ancestor-F1 & Acc & Edge-F1 & Ancestor-F1 & Acc & Edge-F1 & Ancestor-F1 \\
        \midrule
        Random  & 50.01 & 48.24 & 49.18 & 50.09 & 48.32 & 49.13 & 49.91 & 48.14 & 49.23 \\
        KB+Headword & 51.13 & \textcolor{white}{0}4.20 & \textcolor{white}{0}4.20 & 50.90 & \textcolor{white}{0}3.53 & \textcolor{white}{0}3.53 & 50.95 & \textcolor{white}{0}3.71 & \textcolor{white}{0}3.71 \\
        Snowball\cite{agichtein2000snowball} & 53.00 & 11.32 & 11.32 & 56.51 & 23.01 & 23.01 & 53.08 & 18.18 & 18.18 \\
        Substr\cite{bordea2016semeval} & 60.65 & 61.28 & 68.01 & 60.48 & 62.47 & 67.43 & 62.47 & 63.57 & 67.32 \\
        Distance-Parent & 57.46 & 50.94 & 56.64 & 58.00 & 50.39 & 57.71 & 58.61 & 52.65 & 56.31  \\
        Distance-Neighbor & 60.19 & 53.39 & 59.83 & 60.60 & 53.62 & 60.74 & 61.24 & 56.38 & 61.41\\
        Vanilla-BERT & 66.68 & 50.57 & 65.11 & 66.14 & 48.91 & 64.25 & 60.14 & 51.25 & 66.28\\
        TaxoExpan\cite{shen2020taxoexpan}  & 59.54 & 48.26 & 60.96 & 57.34 & 50.45 & 63.48 & 56.36 & 50.79 & 65.41 \\
        TMN\cite{zhang2021taxonomy} & 58.71 & 47.33 & 62.50 & 58.64 & 49.38 & 63.21 & 57.38 & 51.19 & 67.42 \\
        STEAM\cite{yu2020steam} & 60.33 & 53.48 & 68.84 & 60.51 & 54.17 & 71.76 & 61.66 & 54.86 & 70.71\\
        Ours& \textbf{75.64} & \textbf{78.59} & \textbf{83.61} & \textbf{77.58} & \textbf{80.64} & \textbf{88.64} & \textbf{71.53} & \textbf{73.45} & \textbf{80.62} \\
        \bottomrule
    \end{tabular}
    \vspace{-2mm}
\end{table*}

For manual evaluation of accuracy, the result shown in Table \ref{tab:extraction_human} is that the accuracy of query-item concept pairs extracted from user click is around 10\% in three different domains.
Although the proportion of data noise is not tiny, it is worth noting that the coverage of correct hyponymy relations is much more important in this graph construction phase.
Considering that our base of query-item concept pairs is quite large, we can extract abundant potential hyponymy relations from user click logs.
We also observe that the clicked items show a long-tail distribution according to clicked frequency, in \emph{``Bread (面包)''} case, the top five clicked frequency concepts are \emph{``Doughnut (甜甜圈)''} (45), \emph{``Whole-wheat Toast (全麦吐司)''} (29), \emph{``Caterpillar Bread (毛毛虫面包)''} (24), \emph{``Meat Floss Buns (肉松包子)''} (21) and \emph{``Baguette (面包棒)''} (20), which are all correct hyponymy relations.
The noisy concepts are generally concentrated at the end of the distribution.

\subsection{Evaluation on Hypernymy Detection} \label{sec:detection}

\subsubsection{Automatic Evaluation}
Given the three different domain taxonomies mentioned above, we construct the datasets by the self-supervised strategy described in Section 4. We show the detailed statistics of the datasets in Table \ref{tab:dataset}, where the proportion of positive to negative samples is 1:1.
In the positive samples, the proportion of headwords to other words that could be detected for relations is 3:7.
Meanwhile, the proportion of shuffle to replace in the negative samples is 1:1.
Then the datasets are divided into the training set, validation set, and test set in the proportion of 60\%/20\%/20\%.
Then, we conduct a set of experiments on the test datasets to compare our proposed method with other baselines.

In the datasets constructed by the self-supervised strategy mentioned above, we randomly sample 1000 hyponymy relations and manually annotate them for accuracy to demonstrate that the adaptively self-supervised strategy can improve the quality of the datasets.
We find that the accuracy of hyponymy relations detected by headwords in the training set increases to 96.3\%, while the accuracy in the existing taxonomy is 90.1\%.
The hyponymy relations, which can not be detected by headwords, achieve pretty high accuracy (94.8\%) because their source is a professional dictionary in the taxonomy construction phase.

\begin{table*}[!t] 
\caption{Ablation study of relational representation (R) and structural representation ($\text{S}_\text{Random}$ and $\text{S}_\text{C-BERT}$).}
\label{tab:ablation}
\centering
\begin{tabular}{c|ccc|ccc|ccc}   
\toprule
\multirow{2}{*}{\textbf{Representation}} & \multicolumn{3}{c|}{\textbf{Snack}} & \multicolumn{3}{c|}{\textbf{Fruits}} & \multicolumn{3}{c}{\textbf{Prepared Food}}\\
  & Acc & Edge-F1 & Ancestor-F1 & Acc & Edge-F1 & Ancestor-F1 & Acc & Edge-F1 & Ancestor-F1 \\
\midrule
$\text{S}_\text{Random}$ & 57.05 & 50.02 & 67.18 & 58.34 & 56.63 & 71.56 & 57.36 & 52.53 & 64.53 \\
$\text{S}_\text{C-BERT}$ & 57.89 & 50.43 & 67.91 & 58.48 & 57.91 & 72.34 & 58.18 & 53.39 & 65.71\\
R & 63.42 & 63.34 & 76.66 & 63.64 & 60.34 & 76.63 & 64.01 & 61.45 & 68.36 \\
Overall & \textbf{75.64} & \textbf{78.59} & \textbf{83.61} & \textbf{76.35} & \textbf{79.69} & \textbf{84.34} & \textbf{71.53} & \textbf{73.45} & \textbf{80.62} \\
\bottomrule
\end{tabular}
\vspace{-5mm}
\end{table*}

\subsubsection{Manual Evaluation}
Compared with models' performance in self-constructed datasets, verifying their effectiveness in the real scenario is much more important.
To this end, we first employ different methods to extract the appropriate hyponymy relations from query-item pairs in user click logs.
We then count the number of the predicted hyponymy relations (\#Rel) and randomly sample 1000 pairs in the Snack domain from these models, respectively.
Three taxonomists are then asked to annotate these samples to calculate the precision of different models.
The predicted hyponymy relation is correct when two and above taxonomists approve.

\subsubsection{Metrics}
Following the previous solutions\cite{song2021should}, we adopt these metrics as evaluation criteria.
\begin{itemize}
    \item \textbf{Accuracy(Acc)} calculates the predicted edge ($e_{pred} \in \mathcal{E}_{pred}$) exactly matches the ground truth ($e_{gt} \in \mathcal{E}_{gt}$).
    \vspace{-3mm}
    \begin{equation}
    \vspace{-2mm}
    \small
        \text{Acc} = \frac{1}{k}\sum_{i=1}^{k}{\mathbb{I}\left(e_{pred}={e}_{gt}\right)}.
    \end{equation}
    \item \textbf{Edge-F1} measures based on the model predicted edges ($\mathcal{E}_{pred}$) and golden edges ($\mathcal{E}_{gt}$).
    \begin{equation}
    \small
        \text{P} = \frac{|\mathcal{E}_{pred} \cap \mathcal{E}_{gt}|}{|\mathcal{E}_{pred}|},
        \quad
        \text{R} = \frac{|\mathcal{E}_{pred} \cap \mathcal{E}_{gt}|}{|\mathcal{E}_{gt}|}.
    \end{equation}
    \item \textbf{Ancestor-F1} adopts a more relax way because it extends all the ancestor-child edges as ground truth edges ($\mathcal{E}^{*}_{gt}$) of the model predicted edges ($\mathcal{E}_{pred}$).
    \begin{equation}
    \small
        \text{P} = \frac{|\mathcal{E}_{pred} \cap \mathcal{E}^{*}_{gt}|}{|\mathcal{E}_{pred}|},
        \quad
        \text{R} = \frac{|\mathcal{E}_{pred} \cap \mathcal{E}^{*}_{gt}|}{|\mathcal{E}^{*}_{gt}|}.
    \end{equation}
\end{itemize}

\subsubsection{Baselines}
We compare our proposed method with several baselines. It is worth noting that the search space for all the baselines is the query-item concept pairs in user behavioral data as same as ours. On the one hand, we keep this setting the same to ensure fairness. On the other hand, using the whole search space is harmful to the F1-score of all the baselines.
\begin{itemize}
    \item \textbf{Random} randomly attaches new concepts to the existing taxonomy.
    \item \textbf{KB+Headword} indicates that the hyponymy relation between A and B can be retrieved from relational knowledge bases (CNDBpedia\cite{xu2017cn} and CNProbase\cite{chen2019cn}), while A is also the headword of B.
    \item \textbf{Snowball}\cite{agichtein2000snowball} generates patterns and extracts relational tuples from our user-generated content.
    \item \textbf{Substr}\cite{bordea2016semeval} regards A as B's hypernymy if A is a substring of B (e.g., \emph{``Bread (面包)''} and \emph{``Fried Bread (炸面包)''}).
    \item \textbf{Vanilla-BERT} simply adopts vanilla BERT pre-trained with web-scale data to classify the hyponymy relations.
    \item \textbf{Distance-Parent} scores the cosine distance between query concept and item concept. The distance less than the threshold is chosen as a predicted hyponymy relation.
    \item \textbf{Distance-Neighbor} scores the cosine distance with both query concept and its children. The distance less than the threshold is chosen as a predicted hyponymy relation.
    \item \textbf{TaxoExpan}\cite{shen2020taxoexpan} is a self-supervised method for taxonomy expansion.
    Its main contribution is injecting the position information to graph neural networks.
    We adopt BERT embedding for TaxoExpan instead of the word embeddings as in the original paper to a fair comparison.
    \item \textbf{TMN}\cite{zhang2021taxonomy} consists of one primal and multiple auxiliary scorers to discover hyponymy relations.
    \item \textbf{STEAM}\cite{yu2020steam} is a self-supervised method that samples mini-paths from the existing taxonomy and extracts features for query-anchor from multiple views.
\end{itemize}

\subsubsection{Evaluation Results}
We first conduct a set of automatic evaluations on different methods as shown in Table \ref{tab:compare}.
Because we set a proportion of 1:1 to the positive and negative samples, the baseline of random is close to 50\%.
The performances of both KB+Headword and Snowball on accuracy are better than random. However, their Edge-F1 is hugely terrible because their precision is perfect, while their recall is only about 2\% and 10\%, respectively, due to the coverage of general knowledge bases and constrained patterns.
By contrast, Substr performs better, indicating that the headword is an essential feature in determining hyponymy relations.
We observe that introducing children nodes brings remarkable improvement towards distance-based methods.
This phenomenon proves that the children nodes can be considered complements to concepts' semantics.
As for vanilla BERT, the result indicates that BERT performs much better on negative samples than positive samples.
For three state-of-the-art methods, TaxoExpan and TMN, have comparable performances which underperform STEAM on all three datasets.
The major drawback of TaxoExpan is that it only relies on the signal of propagation among neighbors in the taxonomy via graph neural networks, which are not sufficient.
The primal and auxiliary scorers in TMN are limited to extracting various features to model hyponymy relations.
STEAM achieves the most robust overall performance proves the mini-path-based prediction and the multi-view co-training designs are practical.
Our proposed approach outperforms all baselines by a large margin on all three datasets, suggesting that the knowledge of domain relationships and structural information in user behavioral information is necessary.

\begin{table}[!h] 
\vspace{-3mm}
\caption{Manual Evaluation.}
\label{tab:manual_baseline}
\centering
\begin{tabular}{cccccccccc}   
\toprule
\multirow{2}{*}{\textbf{Method}} & \multicolumn{3}{c}{\textbf{\#Rel}} & \multirow{2}{*}{\textbf{Pre}} \\
  & Snack & Fruits & Prepared Food & \\
\midrule
Distance-Neighbor & 74583 & 14964 & 19583 & 80.3\\
TaxoExpan\cite{shen2020taxoexpan} & 1789554 & 80985 & 56981 & 72.3\\
STEAM\cite{yu2020steam} & 132453 & 658432 & 348595 & 76.3\\
Ours & 63819 & 13448 & 17431 & 88.0 \\
\bottomrule
\end{tabular}
\vspace{-2mm}
\end{table}

For manual evaluation, as shown in Table \ref{tab:manual_baseline}, we care about the model precision of different models. 
Because in the real scenario, we mainly collect the positive relations to expand the existing taxonomy while considering both positive and negative relations in self-constructed datasets.
We find that the total number of predicted hyponymy relations of TaxoExpan and STEAM is large. However, their accuracy is much lower, significantly reducing the quality of the existing taxonomy.
Our proposed method has a similar total number to Distance-Neighbor, and about 8 absolute points improve the performance in accuracy.
In conclusion, our method can enlarge the size of real-world product taxonomies from 39,263 to 94,698 with 88\% precision.

\begin{table*}[!t] 
\caption{Ablation study of various design choices.}
\label{tab:abla_repre}
\centering
\begin{tabular}{cl|ccc|ccc|ccc}   
\toprule
\multirow{2}{*}{\textbf{Representation}} & \multirow{2}{*}{\textbf{Model}} & \multicolumn{3}{c|}{\textbf{Snack}} & \multicolumn{3}{c|}{\textbf{Fruits}} & \multicolumn{3}{c}{\textbf{Prepared Food}} \\
 & & Acc & Edge-F1 & Ancestor-F1 & Acc & Edge-F1 & Ancestor-F1 & Acc & Edge-F1 & Ancestor-F1 \\
\midrule
 & Overall & \textbf{75.64} & \textbf{78.59} & \textbf{83.61} & \textbf{76.35} & \textbf{79.69} & \textbf{84.34} & \textbf{71.53} & \textbf{73.45} & \textbf{80.62}  \\
\midrule
\multirow{3}{*}{Relational} & - Template & 75.01 & 77.59 & 82.75 & 75.57 & 78.12 & 82.35 & 70.58 & 72.53 & 79.51  \\
 & - Finetune & 73.45 & 76.58 & 81.34 & 73.53 & 75.12 & 78.53 & 68.53 & 68.89 & 74.24  \\
 & - Concept-level Masking & 63.34 & 58.42 & 65.26 & 68.53 & 61.46 & 69.54 & 62.36 & 57.74 & 66.36  \\
\midrule
\multirow{3}{*}{Structural} & - Edge Attribute & 74.36 & 74.38 & 82.15 & 74.32 & 76.53 & 77.57 & 70.53 & 72.53 & 79.42 \\
 & - User Click Graph & 70.34 & 71.23 & 79.35 & 72.75 & 63.64 & 65.89 & 70.16 & 70.26 & 73.14 \\
 & - Contrastive Learning & 71.59 & 64.12 & 72.43 & 74.73 & 63.24 & 77.57 & 70.48 & 68.12 & 78.25  \\
 & - Position Embedding & 71.93 & 54.72 & 73.84 & 73.02 & 56.42 & 71.59 & 68.19 & 63.57 & 70.92  \\
\bottomrule
\end{tabular}
\vspace{-5mm}
\end{table*}

\subsection{Framework Analysis} \label{sec:analysis}

\subsubsection{Feature Ablation}
We detect the effectiveness of each representation respectively as shown in Table \ref{tab:ablation}.
$\text{S}_\text{Random}$ represents only using structural representation initialized with random embedding, while $\text{S}_\text{C-BERT}$ represents only using structural representation initialized with C-BERT embedding.
The performance of $\text{S}_\text{C-BERT}$ is slightly better than $\text{S}_\text{Random}$, which means that the representations of C-BERT contain helpful information of concepts.
Only using structural representation ($\text{S}_\text{Random}$ and $\text{S}_\text{C-BERT}$) performs not well to identify the hyponymy relations while relational representation (R) works much better.
The former indicates that the structure information in the graph does not suffice, and the latter can show that the pre-trained language model indeed captures in-domain relational knowledge.
However, despite the bad performance of structural representation ($\text{S}_\text{Random}$ and $\text{S}_\text{C-BERT}$), combining relational representation (R) with it can bring 5.90 absolute improvements in Accuracy and 7.33 in Edge-F1.
This phenomenon strongly indicates that the two features can capture different information and are complementary.

\subsubsection{Relational Representation Ablation}
We further conduct a series of ablation studies to detect the contributions of different components in both structural and relational representations as shown in Table \ref{tab:abla_repre}.
We remove only one component and keep others the same to detect their corresponding contribution to controlling variables.
Firstly, we examine the effectiveness of different design choices in our relational representation to capture relational knowledge automatically.
The template does not have a massive effect on the model performance, and we find that it mainly contributes to distinguishing the correct order of two concepts.
This phenomenon partly accounts that the representation in BERT already captures the relational knowledge but neglects the relative position knowledge.
Fine-tuning on training datasets working well for our task means introducing additional relational knowledge in the fine-tuning phase.
Injecting domain knowledge through concept-level masking strategy into a general pretrained language model improves model performance. This phenomenon shows that, on the one hand, there exists a vast knowledge gap between the general domain and specific domain; on the other hand, concept-level is better than token-level to handle the concept-level task.

\begin{table}[t]
    \caption{{The influence of different variants in graph neural network and contrastive learning on Snack dataset.}}
    \label{tab:depth}
    \centering
    \begin{tabular}{ccccc}
    \toprule
        & \multirow{2}{*}{\textbf{Design Choice}} & \multicolumn{3}{c}{\textbf{Snack}} \\
        & & Acc & Edge-F1 & Ancestor-F1 \\
    \midrule
        \multicolumn{5}{c}{\textbf{Graph Neural Network}} \\
    \midrule
        \multirow{2}{*}{\textbf{Neighbors}} & One-hop & \textbf{75.64} & \textbf{78.59} & 83.61  \\
        & Two-hop & 70.18 & 72.41 & \textbf{94.32} \\
    \midrule
        \multirow{3}{*}{\textbf{Aggregation}} & GCN\cite{kipf2016semi} & \textbf{75.64} & \textbf{78.59} & \textbf{83.61} \\
        & GAT\cite{velivckovic2017graph} & 74.16 & 72.72 & 78.59 \\
        & GraphSAGE\cite{hamilton2017inductive} & 72.86 & 71.77 & 76.92 \\
    \midrule
        \multicolumn{5}{c}{\textbf{Contrastive Learning}} \\
    \midrule
        \multirow{6}{*}{\textbf{Negative Rate}}& 0.8 & 74.15 & 73.59 & 82.11\\
        & 1.0 & 73.82 & 74.61 & 80.32\\
        & \textbf{1.2} & \textbf{75.64} & \textbf{78.59} & \textbf{83.61}\\
        & 1.5 & 71.96 & 72.33 & 78.41\\
        & 2.0 & 63.28 & 69.65 & 70.82\\
    \bottomrule
    \end{tabular}
    \vspace{-6mm}
\end{table}

\subsubsection{Structural Representation Ablation}
For the structural representation shown in Table \ref{tab:abla_repre}, it is interesting to note that performance degradation is more pronounced than in the relational representations.
More precisely, edge weights slightly contribute to model performance, demonstrating that higher frequency concepts play a more important role in structural representation.
However, the most likely reason it does not work as expected is that the edge weights are not used as powerful features to facilitate prediction directly but as a weight matrix to propagate the representation.
The information on user behavior contained in the user click graph can complement the lack of structural representation in existing taxonomies.
The pre-training phase of contrastive learning brings a slight improvement in accuracy but a considerable improvement in Edge-F1, which suggests that it helps extract more positive hyponymy relationships.
Meanwhile, expanding the distance between the forward and reverse samples is vital for model differentiation.
Furthermore, considering that there is no relative positional information in the undirected graph in the setting of graphical convolutional networks, the importance of replenishing positional embedding for identifying hyponymy relations is self-evident.

We further explore the influence of different variants in graph neural networks and contrastive learning on the Snack dataset as an example.
As shown in Table \ref{tab:depth}, one-hop relations (i.e., parents and children) is better than two-hop relations (i.e., grandparents and siblings) in both Accuracy and Edge-F1.
The better performance in Ancestor-F1 of two-hop is reasonable because the primary concern metric to our proposed method is Edge-F1, and we especially pull away from the ancestors from anchor nodes.
Two-hop relations reconsider the information from its grandparents and siblings, which boosts Ancestor-F1 but damages the performance in Edge-F1.
For the aggregation function, we attempt two other typical graph neural networks, Graph attention network (GAT)\cite{velivckovic2017graph} and GraphSAGE\cite{hamilton2017inductive}.
Graph attention network performs better than GraphSAGE, indicating that learning the weight by attention mechanism is helpful to predict hyponymy relations.
The graph convolutional network with customized weights calculated from user behavior information achieves the best performance.

The results about contrastive learning in Table \ref{tab:depth} show that with different negative rates, which is the ratio of positive and negative samples, the model achieves the best result when the rate is 1.2.
It is worth noting that the performances as the ratios varying from 0.8 to 2.0 are still better than all the other baselines to prove the robustness of our method.

\begin{table*}[t] 
    \caption{Case Study. \textbf{Bold} in clicked item examples indicate the correct concepts contained in the clicked item names. \cmark and \xmark \ present the results of human evaluations.}
    \label{tab:casestudy}
    \centering
    \begin{tabular}{P{2cm}c|P{4cm}P{5cm}P{4cm}}
    \toprule
       \textbf{Query Concept} & & \textbf{Bread (面包)} & \textbf{Watermelon (西瓜)} & \textbf{Coarse Cereals (杂粮)}  \\
    \midrule
       \textbf{Domain} & & \textbf{Snack} & \textbf{Fruits} & \textbf{Prepared Food} \\
    \midrule
        \makecell{\textbf{Clicked}\\ \textbf{Item}\\ \textbf{Examples}} &
        & \tabincell{P{4cm}}{
        \textbf{Croissant}/bag (\textbf{牛角包}/袋) \\
        Various \textbf{Toasts} (各式\textbf{吐司}) \\
        \textbf{Creamy Bread} (\textbf{奶霜面包}) \\ 
        \textbf{Golden Carrot Cake Bread} - 6 in a bag 
        (\textbf{黄金萝卜饼面包}-6个装)\\
        \textbf{Turkey Black Sesame Bread} - 4 flavors
        (\textbf{土耳其黑芝麻面包}-4个口味)}
        
        & \tabincell{P{5cm}}{Half A-level \textbf{Hainan Sweet Kirin Melon} (海南冰糖麒麟瓜A级-半个)\\ 
        Delicious \textbf{Heart-shaped Watermelon} \\  (好吃的心形西瓜)\\
        \textbf{Peeled and Diced Watermelon} - about 500g per box (去皮切块西瓜-约500克盒)\\
        \textbf{Iced Watermelon} - 750 ml \\ (冰镇西瓜-750ml) }
        
        & \tabincell{P{4cm}}{Coarse Grains \textbf{Oybeans} 250g \\ (杂粮黄豆  250g) \\ Nutritious \textbf{Oatmeal} (营养燕麦片) \\ Bulk \textbf{Adzuki Beans} (散装赤小豆) \\ About half a kilo of Original Roasted \textbf{Sweet Potato} \\ (原味烤山芋一斤左右一个)\\} 
        \\

       \midrule
        \multirow{8}{*}{\makecell{\textbf{Hyponym}\\ \textbf{Relation} \\ \textbf{Prediction}}} & Positive & \tabincell{P{4cm}}{Croissant (牛角包) \cmark \\ Creamy Bread (奶霜面包) \cmark \\ 
        Golden Carrot Cake Bread \\ (黄金萝卜饼面包) \cmark \\
        Turkey Black Sesame Bread \\ (土耳其黑芝麻面包) \xmark \\ }
        
     & \tabincell{P{5cm}}{Kirin Melon (麒麟瓜) \cmark\\ Heart-shaped Watermelon (心形西瓜) \cmark\\ Watermelon Chunks (切块西瓜) \cmark\\ Iced Watermelon (冰镇西瓜) \cmark \\ }

     & \tabincell{P{4cm}}{Soybeans (黄豆) \cmark\\ Oatmeal (燕麦片) \cmark\\ Adzuki Beans (赤小豆) \cmark\\ Sweet Potatoes (山芋) \cmark\\ Rice (大米) \cmark}
     
     \\

 & Negative 
 & \tabincell{P{4cm}}{Big Fried Dumpling (大锅贴) \cmark\\ Lava Dark Chocolate \\ (熔岩黑巧克力) \cmark\\ Sweet Glutinous Rice Balls \\ (甜汤圆) \cmark\\ Turnip Pancakes (萝卜煎饼) \xmark}

 & \tabincell{P{5cm}}{Watermelon Sundae (西瓜圣代) \cmark\\ Cantaloupe Sundae (哈密瓜圣代) \cmark\\ Fruit Sundae (水果圣代) \cmark\\ Lily Crisp (百合酥) \cmark\\ Cold and dressed Sprout (凉拌芽菜) \cmark}
 
 & \tabincell{P{4cm}}{Original Toast Slices \\ (原味吐司片) \cmark\\ Pickled Minced Vegetables \\ (碎米芽菜) \cmark\\ Black Bean Sprouts \\ (黑豆芽) \cmark}\\
    \bottomrule
    \end{tabular}
    \vspace{-5mm}
\end{table*}

\begin{table}[!t]
    \caption{Self-supervised generated dataset statistics.}
    \label{tab:self-dataset}
    \centering
    \begin{tabular}{cccccc}
    \toprule
       \textbf{Method} & \textbf{$|\mathcal{E}_\text{Head}|$} & \textbf{$|\mathcal{E}_\text{Others}|$} & \textbf{$|\mathcal{E}_\text{Train}|$} & \textbf{$|\mathcal{E}_\text{Val}|$} & \textbf{$|\mathcal{E}_\text{Test}|$} \\
    \midrule
        \textbf{Previous} & 28580 & 3626 & 36161 & 12034 & 12073 \\
        \textbf{Ours} & 1554 & 3626 & 6216 & 2072 & 2072 \\
    \bottomrule
    \end{tabular}
\end{table}

\begin{figure}[!t]
    \centering
    \includegraphics[width=0.5\textwidth]{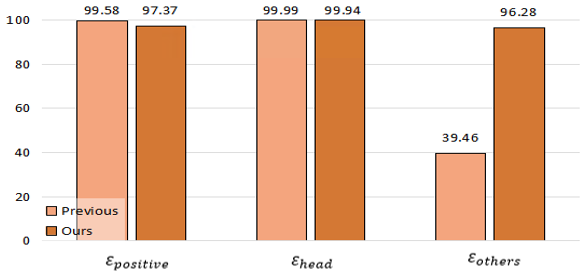}
    \vspace{-5mm}
    \caption{The accuracy of two settings on positive samples.}
    \vspace{-5mm}
    \label{fig:bar}
\end{figure}

\subsubsection{Analysis of Self-supervision}
In this part, we compare the influence of our adaptively self-supervised generation strategy (Ours) to model performance with a self-supervised setting in previous studies (Previous).
We construct datasets from the whole taxonomy in the Snack domain and keep other settings the same as ours.
The statistics of the two datasets are shown in Table \ref{tab:self-dataset}.
Considering our task targets extracting correct hyponymy relations, we analyze the accuracy of two settings on positive samples.
As shown in Figure \ref{fig:bar}, the previous setting achieves higher accuracy than ours on overall positive samples.
However, we observe that this accuracy is inflated because the hyponymy relations, which can be detected by headword, is nearly 100\% but only 39.40\% on others.
Our setting can perform well on both relations with 99.94\% and 96.28\% accuracy, respectively.

\begin{table}[!t]
    \caption{Proportion of Predicted Hyponymy Relations.}
    \label{tab:self-predicte}
    \centering
    \begin{tabular}{cccc}
    \toprule
       \textbf{Method} & \textbf{$\mathcal{E}_\text{All}$} & \textbf{$\mathcal{E}_\text{Head}$} & \textbf{$\mathcal{E}_\text{Others}$}  \\
    \midrule
        \textbf{Previous} & 17385 & 17323 & 61 \\
        \textbf{Ours} & 63819 & 61624 & 2195 \\
    \bottomrule
    \end{tabular}
    \vspace{-5mm}
\end{table}

We further compare the proportion of two types of predicted hyponymy relations from user click logs in both settings as shown in Table \ref{tab:self-predicte}.
Notably, the previous setting predicted fewer total hyponymy relations than ours, and the proportion of other relations to all relations was 0.3\%, much lower than our 3\%.
This phenomenon suggests that the previous setup suffers from data skews and overfitting problems, while our setup can alleviate both problems to some extent.

\subsection{Case Studies}

We can retrieve the corresponding clicked items from the users' click logs given the query concept.
Some examples of clicked items are listed in the second row of Table \ref{tab:casestudy}.
After extracting the item concepts from the clicked items, our model will identify a hyponym relationship between the query concept and each item concept.
The last row of Table \ref{tab:casestudy} shows the prediction results for our selected cases.

The selected cases prove that our model is good at predicting hyponym relations.
Here we take the query concept \emph{``Bread (面包)''} as an example.
Although the concept \emph{``Toast (吐司)''} does not contain the headword \emph{``Bread (面包)''}, our model can still identify the hyponym relationship between them.
However, our model still needs improvement, especially for concepts that do not contain headwords. 
The negative samples in the last row of Table \ref{tab:casestudy} show that our model sometimes incorrectly classifies concepts without headwords as hyponym of query concepts, such as \emph{``Turnip Pancakes (萝卜煎饼)''}.

In addition, the selected cases also show that the classification granularity of our model is not satisfactory enough. 
For example, our model determines that \emph{``Turkey Black Sesame Bread (土耳其黑芝麻面包)''} is a positive sample because it contains the headword \emph{``Bread (面包)''}. 
However, it is more suitable to be regarded as a hyponym of \emph{``Black Sesame Bread (黑芝麻面包)''} rather than \emph{``Bread (面包)''}.

\subsection{Offline User Study of Query Rewriting for Searching}
Following \cite{liu2019user}, we have conducted an offline user study to evaluate how hyponym relations can help improve the search engine through query rewriting.
Our evaluation dataset consists of 100 queries selected from the Meituan take-out application, which is the largest take-out search engine in China.
We rewrite each query $q$ with its hypernym $h$. 
Moreover, we also collect the top 10 search results returned by the Meituan application for each query.

After collecting the search results, we ask three human judges to evaluate the relevance of the search results in the Meituan take-out search engine.
For each search result, we recorded the majority vote of ``relevant'' or ``not relevant'' to calculate the percentage of relevant search results for the original query and the rewritten query, respectively.
Our evaluation results show that the occurrences increase from $74\%$ to $80\%$ after rewriting the query because search engines do not recognize and understand most fine-grained concepts.
To address this issue, our product taxonomy can help provide hypernym queries appropriately to expand the recall of search results and better match the users' intention.

\section{Related Work}
\subsection{Taxonomy Construction}
Recent studies related to taxonomy mainly focus on taxonomy construction from scratch.
According to their differences in the hyponymy detection stage, existing methods can be divided into two categories: 
(1) \emph{pattern-based} methods \cite{hearst1992automatic, kozareva2010semi, nakashole2012patty, panchenko2016taxi, snow2004learning, jiang2017metapad} which leverage lexical-syntactic patterns to extract hyponymy relations between the co-occurrence of term pairs;
(2) \emph{distributional} methods \cite{lin1998information, fu2014learning, rimell2014distributional, shwartz2016improving, luu2016learning, yu2015learning, cocos2018comparing, shen2018hiexpan, bernier2018crim} which calculate pairwise similarity based on term embeddings to predict the hyponymy relations.
Besides, some methods utilize advanced techniques such as transfer learning \cite{shang2020taxonomy}, reinforcement learning \cite{mao2018end} and entity set expansion techniques \cite{shen2017setexpan, zhang2017entity} to adaptively or incrementally construct a taxonomy.

\subsection{Taxonomy Expansion}
Most existing works have designed features to capture abundant representations of concepts. 
Aly \emph{et al.}\cite{aly2019every} adopt hyperbolic embedding to capture hierarchical lexical-semantic relations.
Fauceglia \emph{et al.}\cite{fauceglia2019automatic} use a hybrid method to combine linguistic patterns, semantic web, and neural network for taxonomy expansion.
Manzoor \emph{et al.}\cite{manzoor2020expanding} model the implicit edge semantics to score the hyponymy relevance between node pairs. 
To better maintain the structure of the existing taxonomy, Shen \emph{et al.}\cite{shen2020taxoexpan} propose position-enhanced graph neural networks to encode the relative position of terms and improve the overall quality of taxonomy.
Song \emph{et al.}\cite{song2021should} design a concept sorting model to extract hyponymy relations and sort their insertion order by utilizing the relationship between the newly mined concepts.
Wang \emph{et al.}\cite{wang2021enquire} utilize the hierarchical information of the existing taxonomy by extracting tree-exclusive features in the taxonomy for better taxonomy coherence. 
One limitation of these approaches is that they mainly focus on the general-purpose taxonomies or utilize the general text corpora. Thus, they cannot be easily generalized to specific taxonomies.
Mao \emph{et al.}\cite{mao2020octet} leverage heterogeneous sources of signals such as lexical semantics and structural information to train an end-to-end online catalog taxonomy enrichment model.
Compared with these methods, our proposed method designs relational and structural representations learned from user-generated content and user click logs to model the semantics of in-domain concepts.

Prior studies like \cite{jurgens2016semeval, schlichtkrull2016msejrku} require manual training data to learn features.
Self-supervised constructing training set from existing taxonomy is a common strategy considered to address this problem in recent works\cite{shen2020taxoexpan,mao2020octet,yu2020steam,zhang2021taxonomy}.
In our work, we further propose an adaptively self-supervised generation strategy to avoid inheriting the adverse problems in the existing taxonomy.

\section{Conclusion}
In this paper, we propose an adaptively self-supervised user behavior-oriented product taxonomy expansion framework.
The user behavior information allows our model to learn the domain-specific relational and structural representations while matching the users' intention and cognition.
The adaptively self-supervised generation strategy can construct a high-quality and balanced training dataset that avoids inheriting problems in the existing taxonomy.
Comprehensive experiments conducted on three real-world product taxonomies have shown that the results achieved by our proposed framework improve considerably over state-of-the-art both in automatic and manual evaluations.
We observe that much other information can be incorporated into the model, such as image and merchant information.
In the future, we will further explore other means to determine the problematic cases which need commonsense knowledge or domain knowledge.


\section*{Acknowledgment}

The authors would like to thank anonymous reviews for their helpful comments.
Furthermore, we also thank Juntao Liu, Huaming Wang, and jingping Liu for their insightful discussions; Qianyu He for feedback on related work; Yumeng Shen for annotation; Zongyu Wang, Junjie Fu and Huimin Xu in Meituan team for their supports in experimental data.
This work is supported by National Key Research and Development Project (No. 2020AAA0109302), Shanghai Science and Technology Innovation Action Plan (No.19511120400), and Shanghai Municipal Science and Technology Major Project (No.2021SHZDZX0103).

\balance
\bibliographystyle{IEEEtran}
\bibliography{IEEEabrv,reference}

\end{CJK}
\end{document}